\documentclass{article}

\usepackage{arxiv}
\usepackage{graphicx}
\usepackage[utf8]{inputenc} 
\usepackage[T1]{fontenc}    
\usepackage{hyperref}       
\usepackage{url}            
\usepackage{booktabs}       
\usepackage{amsfonts}       
\usepackage{nicefrac}       
\usepackage{microtype}      
\usepackage{lipsum}
\usepackage{amsmath}
\usepackage{array}
\usepackage{booktabs}
\title{Assessing temporal complementarity between three variable
energy sources by means of correlation and compromise
programming}

\author{
  Fausto A.~Canales \\
    Department of Civil and Environmental \\
  Universidad de la Costa \\ 
   Barranquilla, Colombia\\
   \And
 Jakub~Jurasz\\
  School of Business Society and Engineering\\
  MDH University,\\
Västerås, Sweden \\
  and \\
  Department of Engineering Management \\
AGH University \\
Krakow, Poland \\
   \AND
   Alexandre Beluco \\
   Instituto de Pesquisas Hidráulicas \\
   Universidade Federal do Rio Grande do Sul \\
   Porto Alegre, Brazil
  \And
  Alexander Kies \\
  Frankfurt Institute for Advanced Studies \\
  Goethe University Frankfurt \\
  Frankfurt am Main, Germany
}

\begin{document}
\maketitle

\begin{abstract}
Renewable energies are deployed worldwide to mitigate climate change and push power systems towards sustainability.
However, the weather-dependent nature of renewable energy sources often hinders their integration to national grids.
Combining different sources to profit from beneficial complementarity has often been proposed as a partial solution to
overcome these issues. This paper introduces a novel method for quantifying total temporal energetic complementarity
between three different variable renewable sources, based on well-known mathematical techniques: correlation
coefficients and compromise programming. It has the major advantage of allowing the simultaneous assessment of partial
and total complementarity. The method is employed to study the complementarity of wind, solar and hydro resources on
different temporal scales in a region of Poland. Results show that timescale selection has a determinant impact on the total
temporal complementarity.
\end{abstract}


\section{Introduction}

An increase in global awareness regarding environmental problems and climate change, as well as the constant growth of energy demands all over the world, have facilitated the increase in
recent decades of the fraction of renewable energies in the
power grids and isolated systems in most countries. However,
one of the main concerns regarding some of these energy
sources is their variability, caused by meteorological factors and
usually high and unpredictable fluctuation of the resources,
which frequently poses a challenge for its integration into national power grids \cite{ming2017optimizing}. This unpredictable nature and dependence
on weather conditions and climatic changes constitutes one of
the main drawbacks of stand-alone systems based on a variable
renewable energy sources (VRES), like wind or solar power, because it may result in performance issues, or in the costly oversizing of the system \cite{safaei2015much}.

One of the most feasible options to overcome (partially or
completely) this shortcoming, is to integrate two or more VRES
in a combination in which these sources complement each
other. This is usually referred to as energetic complementarity
and is often expressed in terms of correlation coefficient or a
complementarity index. A complementarity index can be defined as a term used to describe the potential of energy sources
to complement each other on a temporal, spatial or spatiotemporal scale, thereby ensuring supply reliability and minimising
power output fluctuations or shortages. When compared to systems based on a single renewable source, hybrid renewable en-
ergy systems based on the complementarity between two or
more VRES are more likely to improve efficiency and reliability
of the system, whilst reducing energy storage requirements.

For the reasons aforementioned, an appropriate evaluation
of the variability of the available renewable sources and the synergies between them is crucial for a better technical and financial planning, for increasing the integration capacity and maximising social and economic benefits \cite{zhang2018quantitative}.

\subsection{A literature review about energetic complementarity}
Over the last years, many researchers across the world have
conducted investigations related to energetic complementarity
between VRES. To provide some context to the method described in this paper, the next paragraphs present a brief review
of several recent and relevant works about energetic complementarity.

Gburčik et al. \cite{gburvcik2006complementary} presented an analysis of solar and wind energy resources in Serbia, demonstrating that the fluctuation effects of those resources can be reduced by means of energy
storage and complementary use of both sources. These authors
analysed the concept of complementarity from both a spatial
and a temporal perspective; however, their approach did not
apply any universal measure/index. The study conducted by Li
et al. \cite{li2009wind} investigated not only the temporal complementarity of
wind and solar resources (based on correlation coefficient) in a
coastal region of Australia, but also assessed if the availability
of those resources matched the load demand, concluding that
combined operation was enough to supply peak load. A paper by Stoyanov et al. \cite{stoyanov2010} investigated the variations and fluctuations
of solar and wind energy source in eight locations in Bulgaria
and compared it to the temporal electric load distribution in order to observe if production was able to match power consumption. Hoicka and Rowlands \cite{hoicka2011solar} analysed the solar and wind resources complementarity in Ontario, Canada. From their results
they concluded that the combination of solar and wind smoothens the power production curve, when compared to production
from a single source.

Brazil is one of the countries with most studies related to
energetic complementarity. Beluco et al. \cite{beluco2012method}, presented a method
to analyse the impact of energetic complementarity on the performance of hybrid solar-hydro power stations, with a follow-up study assessing several different configurations and re
sources availability \cite{beluco2013influence}. The research by De Jong et al. \cite{de2013solar} investigated the complementarity between solar, wind and hydro resources in the northeast of Brazil, and their combined capacity
to supply peak loads. In their analysis, these authors used the
Pearson correlation coefficient as a complementarity indicator,
and their findings indicate that wind energy has a significant potential to reduce the operation of hydropower during the irrigation period. Ramos et al. \cite{ramos2013minimizing} proposed a model for assessing wind
and hydropower complementarity and its effect on the financial
performance and risk exposures, using as case study 10 locations spread over the country. Silva et al. \cite{silva2016complementarity} analysed the complementarity of Brazilian hydropower with the offshore wind,
by means of comparing monthly precipitation and wind data,
whereas Schmidt et al. \cite{schmidt2016optimal} presented a simulation model for optimising the mix and operation of solar, wind and hydropower
generation in a low-carbon Brazilian power system, utilising the
underlying assumption of resources complementarity. Cantão
et al. \cite{cantao2017evaluation} proposed so-called correlation maps to evaluate the hydro-wind complementarity for the entire Brazilian territory,
based on Pearson and Spearman’s rank coefficients, which were
also used by Denault et al. \cite{denault2009complementarity} . The study by Pianezzola et al. \cite{pianezzola2017complementarity},
based on the method proposed by Beluco et al. \cite{beluco2008dimensionless} , created spatiotemporal complementarity maps for solar and wind resources for the state of Rio Grande do Sul, in southern Brazil.
Rosa et al. \cite{de2017complementarity} used the Pearson correlation coefficient for assessing the complementarity between pairs of hydropower
plants, photovoltaic (PV) stations and wind farms in Rio de
Janeiro. The Pearson correlation coefficient was also used by
Bagatini et al. \cite{bagatini2017complementarity}, for assessing complementarity between hydro,
wind and solar energy in Rio Grande do Sul (Brazil), and their
approach was also based on comparing individual pairs of resources. Recently, Risso et al. \cite{risso2018complementarity}, suggested a concept of complementarity roses to assess the spatial complementarity between
renewable resources and express this feature as maps.

In other complementarity studies around the world, Monforti et al. \cite{monforti2014assessing}, used a Monte Carlo approach to assess the complementarity of solar and wind resources in Italy, obtaining relatively high monthly correlation coefficients. Kunwar \cite{kunwar2014complementarity} assessed
the potential of wind and solar complementarity for compensating the reduced hydropower potential during dry seasons
with low stream flow in rivers used for power generation. Francois et al. \cite{francois2016complementarity} conducted a complementarity analysis for solar and
hydropower resources at northern Italy, considering different
temporal scales and using energy balance and storage requirements as indicators. An interesting approach for increasing
complementarity between small hydropower stations and solar
PV systems was presented by Kougias et al. \cite{kougias2016methodology}. Their method is
based on the suboptimal orientation of PV panels, which sacrifices PV generation, but increases complementarity.

Solomon et al. \cite{solomon2016investigating} used load and VRES data from California,
for investigating the beneficial impact of solar and wind resources temporal complementarity on the storage requirements and resulting reliability of the power system. François et
al. \cite{franccois2017assessing} analysed the solar/hydro complementarity considering the
effect of hydrological uncertainty in ungauged watersheds,
evaluating the method using data from mountain basins in the
Eastern Italian Alps. Jurasz et al. \cite{jurasz2016assessing} used Pearson correlation coefficient to assess the complementarity between solar, wind
and hydrokinetic energies in several locations in Poland.

Xu et al. \cite{xu2017spatial} analysed and mapped the spatiotemporal variations of complementary solar and wind resources over the entire Chinese territory, using the Kendall’s rank correlation coefficient as regionalisation indicator. Gulagi et al. \cite{gulagi2017solar} investigate how a combination of solar-wind complementarity, storage and
transmission can be used to mitigate the monsoon effects in India. Prasad et al. \cite{prasad2017assessment} analysed the effect of spatial and temporal
synergy between solar and wind resources in Australia, and presented a method that can be readily applied in other parts of
the world to investigate synergies between these two VRES.
During Filho et al. \cite{fo2018influence} analysed how temporal complementarity
affects storage requirements in solar-hydro hybrid systems.

Jurasz et al. \cite{jurasz2017temporal} analysed the complementarity of solar and
wind energy over Lower Silesia in Poland. In their follow-up
study \cite{jurasz2018large} they presented how complementarity between solar
and wind resources, coupled with pumped storage hydropower,
can be used from the perspective of guaranteeing power system reliability. Jurasz et al \cite{jurasz2018impact} investigated how the theoretical
complementarity between solar and wind resources can be
used to assess the hybrid system reliability, and the impact of
various levels of energy storage on the system’s overall performance.

More recently, energetic complementarity has also been
employed in the formulation of optimisation problems. Jurasz
and Mikulik \cite{jurasz2017site} presented a simple mathematical model for selecting an optimal location for solar and wind parks, aiming to
reduce the ramp rates of aggregated solar-wind generation, using data from several locations in Poland, considering VRES
complementarity on a temporal and spatial scale. Aza-Gnandji
et al. \cite{aza2018complementarity} assessed the complementarity between solar and wind
energy in Benin Republic. In their study they determined the
best geographical locations in terms of complementarity by
means of Particle Swarm Optimisation. Zhu et al. \cite{zhu2018dynamic} proposed an
economic dispatch strategy based on multi-scale complementary energy sources. These authors proposed an optimisation
model, whose objective function was minimising the required
output from thermal plants, using as case study data from a
power system in southeast China. Zhu et al. \cite{zhu2018complementary} presented a complementary operation for hydro-wind-solar system in Jinsha
River. In their paper, the authors evaluated the complementarity of the different paired combinations of VRES and set up an
optimisation model whose objective was minimising the hydro-
power output. Using as case study the wind, solar and hydro-
power resources in the Yalong River basin and their output
complementarity, Zhang et al. \cite{zhang2018short} formulated an optimisation
model with the objective of minimising the excess energy from
wind farms and photovoltaic installations, whilst maximising
the stored energy in cascade hydropower stations. Shaner et
al. \cite{shaner2018geophysical} analysed the impact of geophysical variability of solar and
wind resources on the system’s reliability, storage requirements
and transmission infrastructure, testing different mixes of these
two sources. Henao et al. \cite{henao2019optimising} presented an optimisation model to
increase the penetration of complementary VRES into the Colombian power grid, aiming at minimising system costs, greenhouse gases emissions and blackout events.

Some authors have investigated energetic complementarity
on geographical scales larger than countries. Francois et al. \cite{franccois2016increasing}
proposed to utilize run-of-river hydro power to increase the
share of VRES in the energy mix, using as case study 12 European regions. Krutova et al. \cite{krutova2017smoothing} analysed the smoothing effect of
VRES in the context of a supergrid covering Eurasia and Africa,
and their findings suggest that the complementarity between
the renewable resources of the region could heavily reduce
backup energy requirements. Miglietta et al. \cite{miglietta2017local} assessed the local complementarity of wind and solar resources over Europe
from a meteorological perspective, using a time series comprising data from three complete years. Sterl et al. \cite{sterl2018new}, based on a
case study from West Africa, proposed a new metric called the
stability coefficient, created for assessing the solar-wind synergies and limiting energy storage needs. Complementarity can
also be achieved within single renewable resources. This is
sometimes referred to as system-friendliness and can include
wind turbines suitable for low wind situations \cite{hirth2016system} or solar PV modules with orientations chosen to resemble the load pattern \cite{chattopadhyay2017impact}.

Complementarity is generally expressed in terms of correlation coefficients or complementarity indices. In most cases,
these metrics are used for measuring complementarity between two VRES. However, a few authors have extended the
existing methods in order to assess energetic complementarity
between more than two sources. Borba and Brito \cite{borba2017index} , extending
on the method presented by Beluco et al. \cite{beluco2008dimensionless}, proposed a dimensionless index for calculating temporal complementarity between two or more energy resources. A recent and interesting
study by Han et al. \cite{han2019quantitative} evaluates the complementarity degree between wind, solar and hydropower by means of comparing the
fluctuation difference between their individual and combined
power generation capacities.
\subsection{Novelty, objective and structure of the paper}
The previous literature review allows noticing some common
features of the research on VRES complementarity:
\begin{itemize}
    \item the Pearson correlation coefficient is the most commonly applied metric used for energetic complementarity assessment,
although some authors follow different approaches, especially in optimisation models proposed for improving the
power system operation based on complementarity;
\item complementarity is mostly investigated for a combination of
only two renewable sources for a given location;
\item the analysis on complementarity is performed in three ways:
it is either temporal, spatial or spatiotemporal;
\item some authors are aiming at analysing or optimising the balance between VRES power output and load demand besides
investigating the resources complementarity;
\item the complementarity between resources is mainly used to increase system reliability, to boost the economic operation and
to improve dispatching strategies;
\item the existing approaches for calculating complementarity indices for more than two sources are essentially unidimensional.
Thus, information regarding complementarity between a specific paired combination of resources becomes unavailable to
the reader in such methods, but on the other hand, correlation coefficient tables do not directly provide information
about the total complementarity between the set of resources.
\end{itemize}
Within this context, the objective of this paper is to introduce a method that simultaneously assesses partial and total
temporal complementarity between three variable energy
sources, using a combination of correlation coefficients, Euclidean vectors, compromise programming and normalisation.
Some benefits and novelties of this approach are the following:
\begin{itemize}
    \item The mathematical techniques included in the method are well
known and ensure the linearity of the results, facilitating their
understanding.
\item Correlation coefficients are employed for estimating complementarity between each pair of resources, and the Euclidean
vector approach allows a bundle representation.
\item Compromise programming and normalisation allows to quantify the total temporal complementarity, ranging from total
similarity (worst-case scenario) to a maximum feasible complementarity (best-case scenario).
\item By calculating correlation coefficients, compromise programming and normalisation, the method is versatile enough for
assessing complementarity between time series, power output, etc. in different timescales.
\end{itemize}
The rest of the paper has the following structure: Section 2
introduces and describes the method for evaluating energetic
complementarity between three VRES by means of correlation
coefficients, a vector representation and compromise programming; Section 3 presents and discusses a case study that evalu-
ates energetic complementarity between three VRES on three
different timescales, using hourly information from a southern
Poland region as source data; finally, Section 4 provides a short
summary of the paper as well as some concluding remarks and
prospective researches derived from this work.

\section{Method}
The method presented in this paper can be used for determining the total temporal complementarity between three variable
energy sources, by means of a combination of proven mathematical techniques. The following subsections describe the
steps of the method.
\subsection{Correlation between each pair of resources and vector representation}
\begin{figure}
    \centering
    \includegraphics[width=\textwidth]{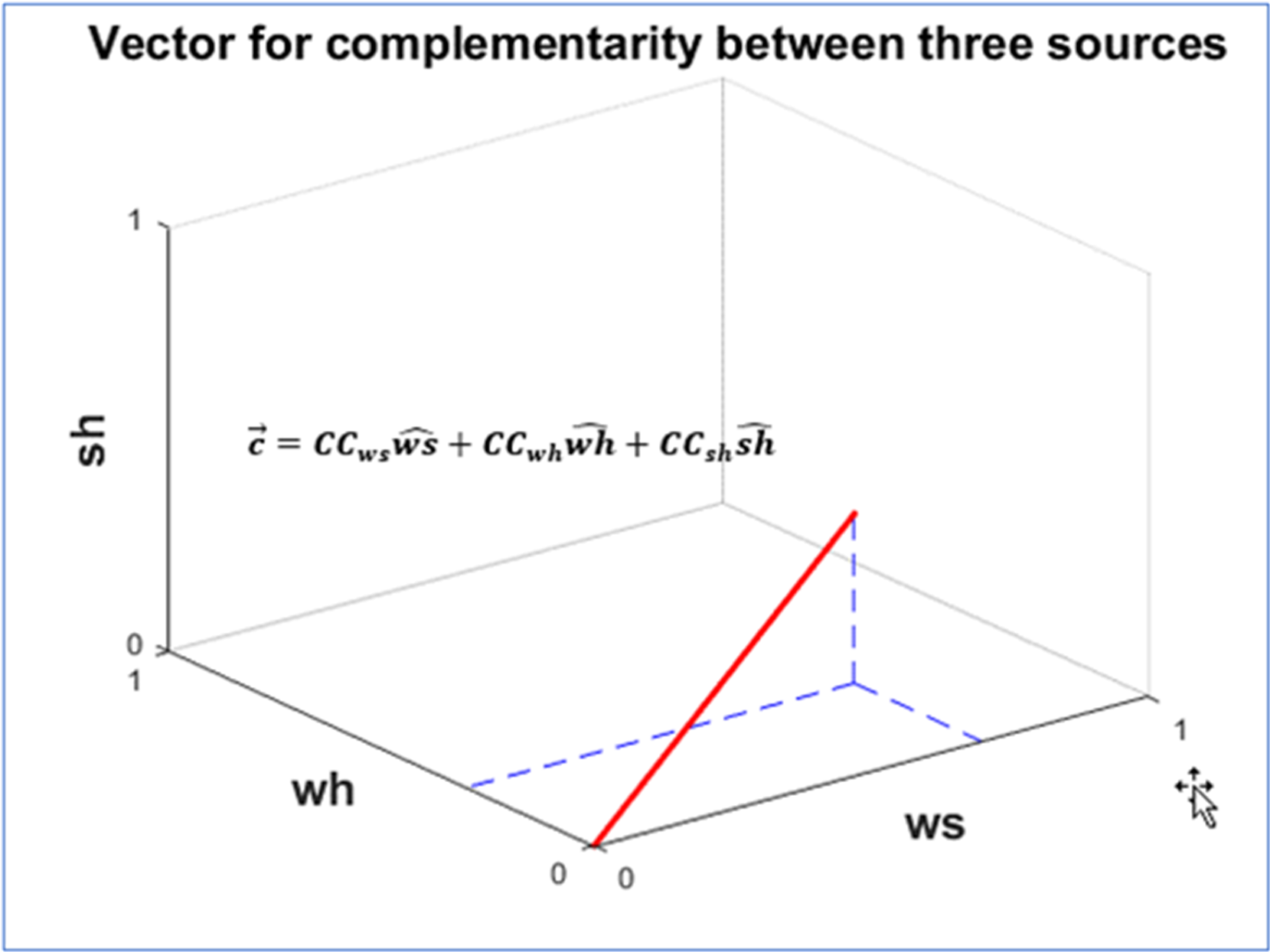}
    \caption{Generic representation of a vector for describing complementarity between three energy sources.}
    \label{fig:1}
\end{figure}
Correlations are measurements of the degree of dependence
between two random variables. The most common type of correlation is the Pearson correlation coefficient $\rho_{xy}$, which is widely
used in energetic complementarity studies. For a sample of
paired data {$(x_1 ,y_1 ),...,(x_n ,y_n)$}, $\rho_{xy}$ can be calculated as:
\begin{align}
    \rho_{xy} &= \frac{\text{Cov}(x,y)}{\rho_x\rho_y} \\
    &=\frac{\sum_{i=1}^n(x_i - \bar{x})(y_i - \bar{y})}{\sqrt{\sum_{i=1}^n(x_i-\bar{x})^2}\sqrt{\sum_{i=1}^n(y_i-\bar{y})^2}} \nonumber
\end{align}
where n is the size of the sample, $x_i$ and $y_i$ are the individual
sample points of each variable, $\bar{x}$ and $\bar{»}$ are the averages for
each sample, $\sigma_x$ and $\sigma_y$ are the standard deviations. However,
for non-normal multivariate distributions, the Pearson correlation coefficient is not a desirable measure of association \cite{denault2009complementarity}.
For non-normal multivariate distributions, Spearman’s rank
correlation coefficient $\rho_{xy}$ is a better alternative for assessing
the dependence between two random variables, because it removes the relative size of the two variables, and the dependence is measured between the transformed variables \cite{carmona2014statistical}. For a
sample of size n, the calculation of $\rho_{xy}$ needs the conversion
of each $x_i$ and $y_i$ value to the corresponding rank $x_i$, and rank $y_i$ :
\begin{align}
    \rho_s(x,y) &= \frac{\text{Cov}(\text{rank x}, \text{rank y)}}{\sigma_{\text{rank x}}\sigma_{\text{rank y}}}
\end{align}
where $\sigma_{\text{rank x}}$ and $\sigma_{\text{rank y}}$ are the standard deviations of the corresponding ranks.
The values of these two types of correlation coefficients will
always be within the interval [-1, 1]. Both Pearson and Spearman’s coefficients are evaluated in this paper for assessing energetic complementarity. According to Cantão et al. \cite{cantao2017evaluation}, one possible interpretation of correlation coefficient values regarding
energetic complementarity is presented in Table 1. A correlation coefficient value close to 0 implies that there’s essentially
no association between the two variables; a positive value indicates that as the value of one of the variables increases or decreases, the value of the other variable follows a similar behaviour; on the other hand, a negative value means that as the
value of one variable increases, the value of the other variable
decreases, and vice versa \cite{vega2017evaluation}.
For this method, each correlation coefficient (CC) found in
the previous step becomes one component of a complementarity vector in a multidimensional space, where each dimension
is defined by a paired combination of energy sources, and the
CC value represents the similarity or complementarity between
each pair of resources. For example, let subindices w, s and h
represent wind, solar and hydropower resources, correspondingly. These resources allow three possible paired combinations, and the resulting three-dimensional vector c, represent-
ing the temporal complementarity between the three energy
sources can be specified in the following form:
\begin{align}
    c = CC_{ws}\hat{ws} + CC_{wh}\hat{wh} + CC_{sh}\bar{sh}
\end{align}
If we considered only one octant of the three-dimensional
space of this general example, a generic representation of a
complementarity vector is shown in Figure 1.
\begin{table}
    \centering
    \begin{tabular}{cccc}
    \hline
       Behaviour  & Correlation coefficient values & Normalisation of correlation coefficient & Interpretation \\\hline
        Similarity & $0.9 \leq CC \leq 1.0$ &$0.00 \leq$ Norm. (CC) $< 0.05$& Very strong similarity \\
        & $0.6 \leq CC \leq 0.9$ &$0.05 \leq$ Norm. (CC) $< 0.20$& Strong similarity \\
        & $0.3 \leq CC \leq 0.6$ &$0.20 \leq$ Norm. (CC) $< 0.35$& Moderate similarity \\
        & $0.0 \leq CC \leq 0.3$ &$0.35 \leq$ Norm. (CC) $< 0.50$& Weak similarity \\
        \hline
        Complementarity & $-0.3 \leq CC \leq 0.0$ &$0.50 \leq$ Norm. (CC) $< 0.65$& Weak complementarity \\
        & $-0.6 \leq CC \leq 0.3$ &$0.65 \leq$ Norm. (CC) $< 0.80$& Moderate complementarity \\
        & $-0.9 \leq CC \leq 0.6$ &$0.80 \leq$ Norm. (CC) $< 0.95$& Strong complementarity \\
        & $-1.9 \leq CC \leq 0.9$ &$0.95 \leq$ Norm. (CC) $< 1.00$& Very strong complementarity \\\hline
    \end{tabular}
    \caption{Interpretation of correlation coefficient values (adapted from \cite{cantao2017evaluation}).}
    \label{tab:my_label}
\end{table}
\subsection{Compromise programming}
Compromise programming is a multi-criteria analysis technique
that focuses on finding the closest point to the ideal solution,
within the domain of the feasible solutions. According to Gershon and Duckstein \cite{gershon1983multiobjective} the metric $L_p$ used in compromise programming for estimating the distance of each option x to the
optimal solution (usually unfeasible) can be found by means of
the following distance function:
\begin{align}
    L_p(c) = \left[\sum_{k=1}^n \alpha_k^p \left|\frac{f_k^\text{best} - f_k(c)}{f_k^\text{best} - f_k^\text{worst}}\right|^p\right]^{\frac{1}{p}}
\end{align}
where: $\alpha_k^p$ are the weights for each component k (where k is
each paired combination). The method presented in this paper
considers that all paired combinations have the same importance, therefore, $\alpha_k^p = 1$ for all cases. Also in equation (3),
$f_k(c)$ is the CC value for the corresponding paired combination
of resources of vector c; $f_k^\text{best}$ is the most desirable value of the
correlation functions, therefore, $f_k^\text{best} = -1$, because it would
represent full complementarity; $f_k^\text{worst}$ is the less desirable value
of the correlation functions, therefore, $f_k^\text{worst} = 1$, because it
would represent full similarity (i.e., the simultaneous occurrence of the resources); $p$ is the parameter that establishes the
type of geometrical distance between $f_k^\text{best}$ and $f_k(c)$. As explained by Gershon and Duckstein \cite{gershon1983multiobjective} , for p = 1, all deviations
from $f_k^\text{best}$ are considered in direct proportion to their magnitudes. For 2 (Euclidean distance) $\leq p < \infty$, the largest deviation
has the greatest influence. The present method adopts the
value of p = 1, allowing a linear assessment of complementarity.
\subsection{Total temporal complementarity index}
It can be easily observed from equation (3) that the corresponding minimum and maximum values of $L_p(c)$ are 0 (perfect complementarity for all paired combinations) and $n^\frac{1}{p}$ (perfect similarity for all paired combinations). However, for the conditions
previously defined and three energy sources, results from heuristics and linear programming indicate that the minimum
achievable Lp(c) equals 0.75 (one example is when every correlation coefficient for each paired combination equals -0.5). A
proof of this is given in the appendix. Therefore, it is possible
to state that a moderate correlation between two VRES does not
mean is impossible to achieve a system 100\% based on renewable energy resources, because a third source could fill the deficits of the other two, thus minimising backup power requirements.

A total temporal complementarity index $\kappa_t$ can be assessed
by normalising the $L_p(c)$ metric through the following expression:
\begin{align}
    \kappa_t(c) = \frac{3-L_p(c)}{2.25}
\end{align}
with $\kappa_t$ values ranging from 0 (perfect similarity) to 1 (perfect
complementarity).

\section{Case study and discussion of results}
For a better understanding of the method presented in this paper, this section presents and discusses a case study that evaluates the complementarity between three variable energy
sources (wind speed, solar irradiation and stream flow rate – or
discharge –) on three different timescales (monthly, daily and
hourly). The dataset used in this work corresponds to the hourly
measurements of the resources for the 8784 hours of 2008.
\subsection{Input data}
\begin{figure}
    \centering
    \includegraphics[width=\textwidth]{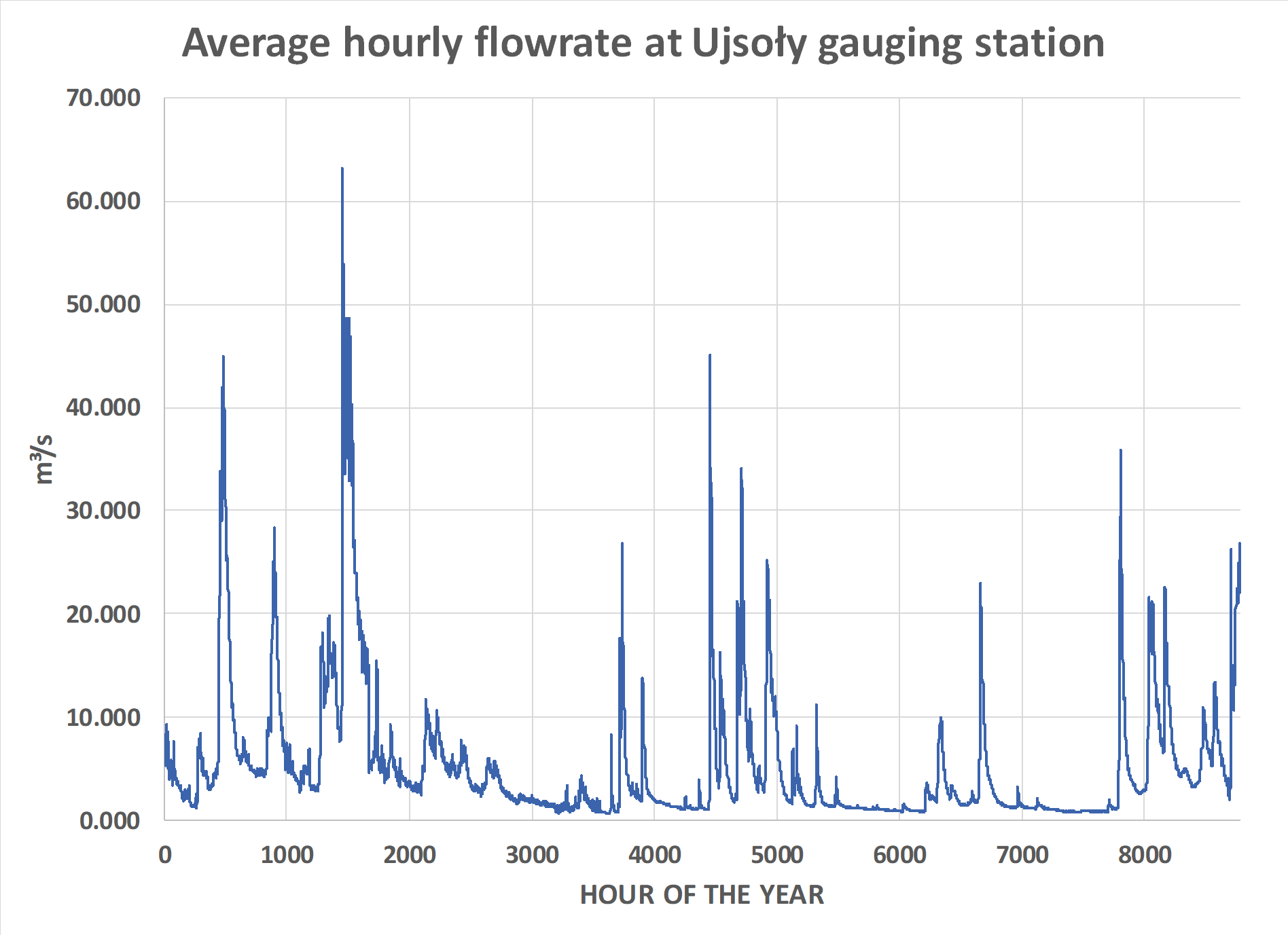}
    \caption{Average hourly flow rate at Ujsoły gauging station for the
year 2008.}
    \label{fig:2}
\end{figure}
\begin{figure}
    \centering
    \includegraphics[width=\textwidth]{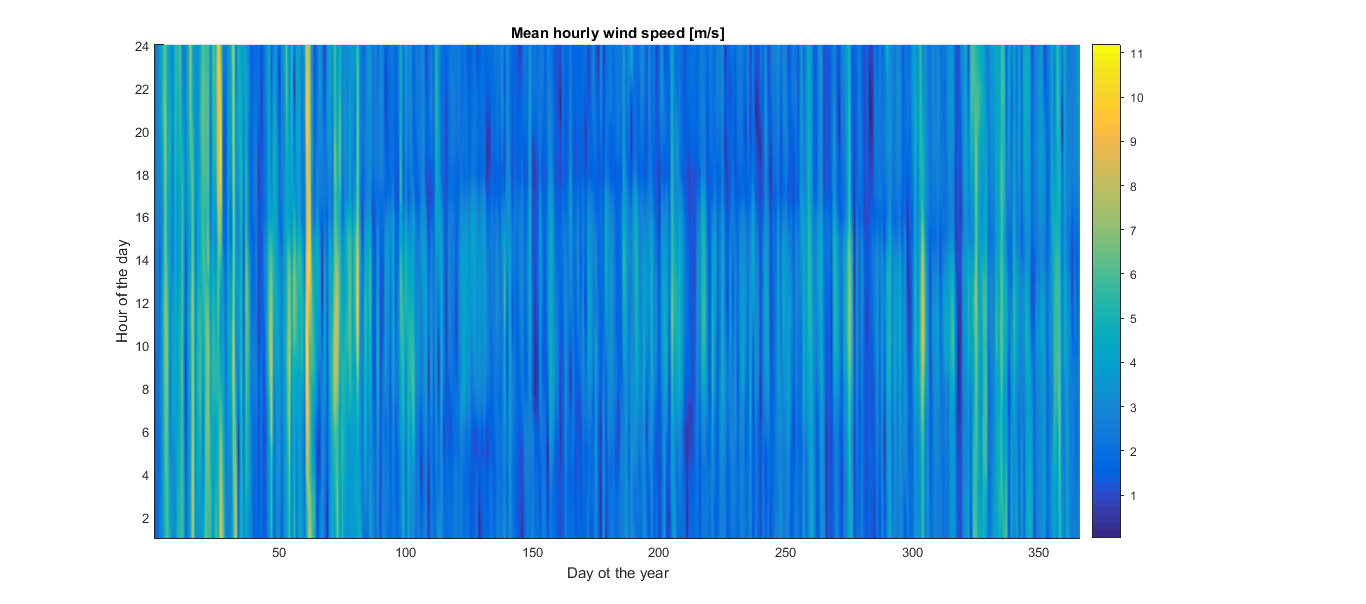}
    \caption{Data map for mean hourly wind speed at the case study
area.}
    \label{fig:3}
\end{figure}
\begin{figure}
    \centering
    \includegraphics[width=\textwidth]{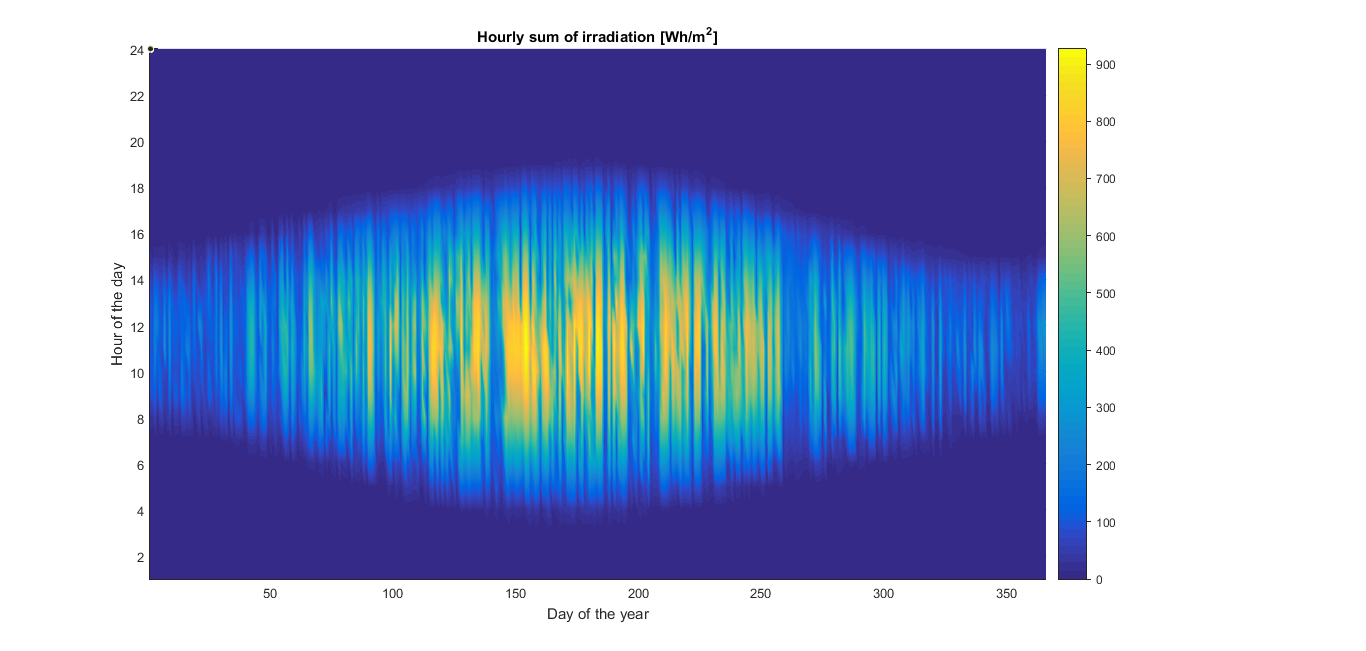}
    \caption{Data map for hourly solar irradiation at the case study
area.}
    \label{fig:4}
\end{figure}
The study area is situated in southern Poland, within the Soła
river basin. At the end of its 89 km course, the Soła river has a
mean discharge of 19.3 $\frac{m^3}{s}$, and a watershed covering approximately 1391 km$^2$ \cite{helios2005monitoring}. For our case study, the stream flow data
corresponds to the discharge measured at the Ujsoły gauging
station, located at coordinates 49$^\circ$29'33"N 19$^\circ$07'01"E, in a region mainly covered by forests and agropastoral farming. The
discharge time series for the Soła river at the Ujsoły gauging station was obtained from the site of the Polish Institute of Meteorology and Water Management - National Research Institute
(IMGW-PIB) at: https://danepubliczne.imgw.pl. The hydrograph
corresponding to the average hourly flow rate for 2008 is shown
in Figure 2.

Wind speed and solar irradiation time series for this area
correspond to satellite measurements \cite{54,55} . It is worth noticing
that previous studies in the area have demonstrated a good fit
between this records and ground data \cite{jurasz2017temporal}. The data maps describing the behaviour of these two resources along the 8784 hours
of 2008 are shown in Figure 3 (wind speed) and Figure 4 (solar
irradiation).
\subsection{Results and discussion}
\begin{table}
    \centering
    \begin{tabular}{c|cccccccccccc}
       \hline Month & Jan & Feb & Mar & Apr & May&Jun&Jul&Aug&Sep&Oct&Nov &Dec  \\\hline
   wind speed [m/s]    &4.402&3.646&3.945&2.656&2.297&2.151&2.585&2.339&2.681&2.815&3.408&3.298  \\
      solar irradiation [$\frac{kW}{m^2 d}$]  &0.768&1.467&2.681&3.825&5.077&5.895&4.992&4.779&3.027&1.952&0.984&0.712  \\
     river flow rate [$\frac{m^3}{s}]$   &7.911&8.619&12.454&4.491&1.625&2.911&8.000&1.897&2.104&2.501&3.212&8.936 \\\hline
     \multicolumn{13}{c}{Pearson correlation coefficient - $\rho_{xy}$} \\\hline
     \multicolumn{6}{c}{Complementarity vector}&\multicolumn{7}{c}{- 0.815 ws + 0.717 wh - 0.410 sh} \\
     \multicolumn{6}{c}{Compromise programming}&\multicolumn{7}{c}{1.246} \\
     \multicolumn{6}{c}{Total temporal complementarity index - $\kappa_t$}&\multicolumn{7}{c}{77.96\%} \\\hline
     \multicolumn{13}{c}{Spearman's rank correlation coefficient - $\rho_s$} \\\hline 
          \multicolumn{6}{c}{Complementarity vector}&\multicolumn{7}{c}{- 0.867 ws + 0.650 wh - 0.517 sh} \\
     \multicolumn{6}{c}{Compromise programming}&\multicolumn{7}{c}{1.133} \\
     \multicolumn{6}{c}{Total temporal complementarity index - $\kappa_t$}&\multicolumn{7}{c}{82.98\%} \\\hline
    \end{tabular}
    \caption{Data and complementarity results on a monthly scale for the case study.}
    \label{tab:2}
\end{table}
\begin{table}
    \centering
    \begin{tabular}{c|c}
    \hline\multicolumn{2}{c}{Pearson correlation coefficient - $\rho_{xy}$} \\\hline
     \multicolumn{1}{c}{Complementarity vector}&\multicolumn{1}{c}{- 0.441 ws + 0.360 wh - 0.287 sh} \\
     \multicolumn{1}{c}{Compromise programming}&\multicolumn{1}{c}{1.316} \\
     \multicolumn{1}{c}{Total temporal complementarity index - $\kappa_t$}&\multicolumn{1}{c}{74.80\%} \\\hline
     \multicolumn{2}{c}{Spearman's rank correlation coefficient - $\rho_s$} \\\hline 
          \multicolumn{1}{c}{Complementarity vector}&\multicolumn{1}{c}{- 0.450 ws + 0.312 wh - 0.356 sh} \\
     \multicolumn{1}{c}{Compromise programming}&\multicolumn{1}{c}{1.253} \\
     \multicolumn{1}{c}{Total temporal complementarity index - $\kappa_t$}&\multicolumn{1}{c}{77.60\%} \\\hline
    \end{tabular}
    \caption{Complementarity results on a daily scale for the case study.}
    \label{tab:3}
\end{table}
\begin{table}
    \centering
    \begin{tabular}{c|c}
    \hline\multicolumn{2}{c}{Pearson correlation coefficient - $\rho_{xy}$} \\\hline
     \multicolumn{1}{c}{Complementarity vector}&\multicolumn{1}{c}{- 0.017 ws + 0.294 wh - 0.132 sh} \\
     \multicolumn{1}{c}{Compromise programming}&\multicolumn{1}{c}{1.573} \\
     \multicolumn{1}{c}{Total temporal complementarity index - $\kappa_t$}&\multicolumn{1}{c}{63.40\%} \\\hline
     \multicolumn{2}{c}{Spearman's rank correlation coefficient - $\rho_s$} \\\hline 
          \multicolumn{1}{c}{Complementarity vector}&\multicolumn{1}{c}{+0.048 ws + 0.244 wh - 0.118 sh} \\
     \multicolumn{1}{c}{Compromise programming}&\multicolumn{1}{c}{1.587} \\
     \multicolumn{1}{c}{Total temporal complementarity index - $\kappa_t$}&\multicolumn{1}{c}{62.80\%} \\\hline
    \end{tabular}
    \caption{Complementarity results on an hourly scale for the case study.}
    \label{tab:4}
\end{table}
\begin{figure}
    \centering
    \includegraphics[width=\textwidth]{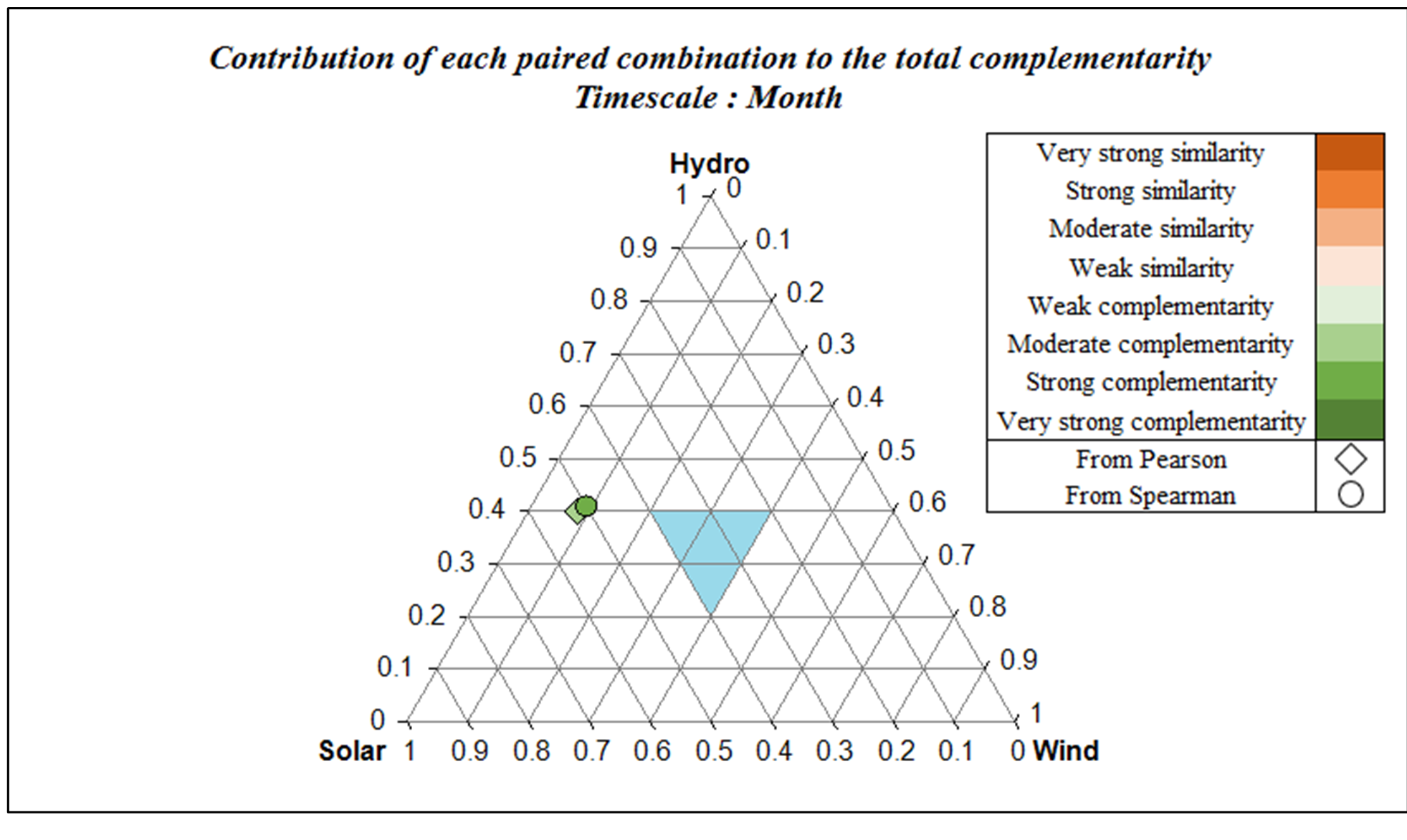}
    \caption{Contribution of each paired combination of VRES to $\kappa_t$ on
a monthly scale.}
    \label{fig:5}
\end{figure}
\begin{figure}
    \centering
    \includegraphics[width=\textwidth]{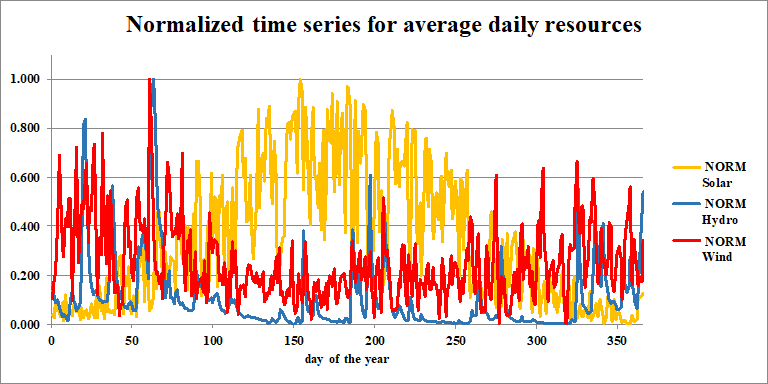}
    \caption{Normalized time series for average daily resources.}
    \label{fig:6}
\end{figure}
\begin{figure}
    \centering
    \includegraphics[width=\textwidth]{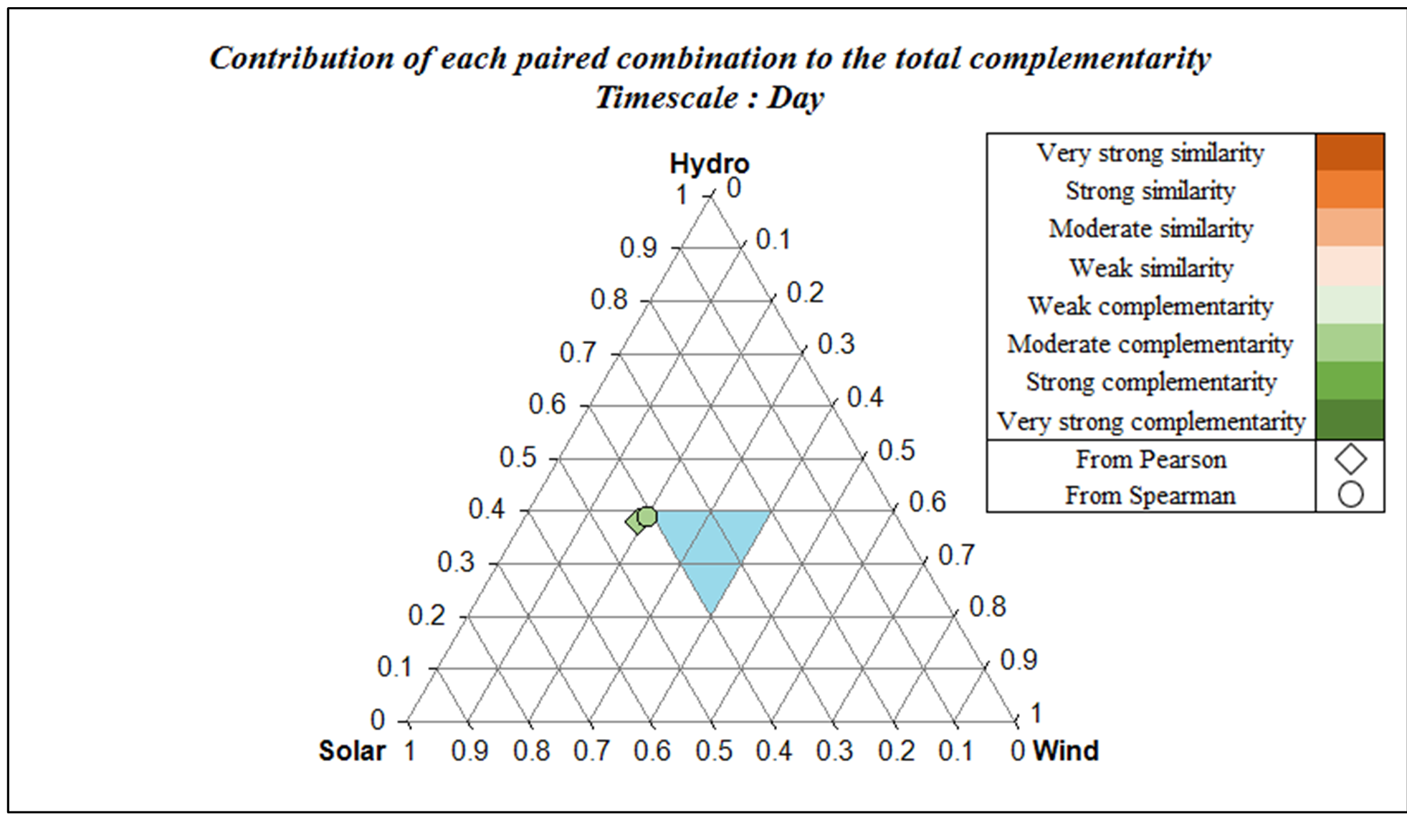}
    \caption{Contribution of each paired combination of VRES to $\kappa_t$ on a
daily scale.}
    \label{fig:7}
\end{figure}
\begin{figure}
    \centering
    \includegraphics[width=\textwidth]{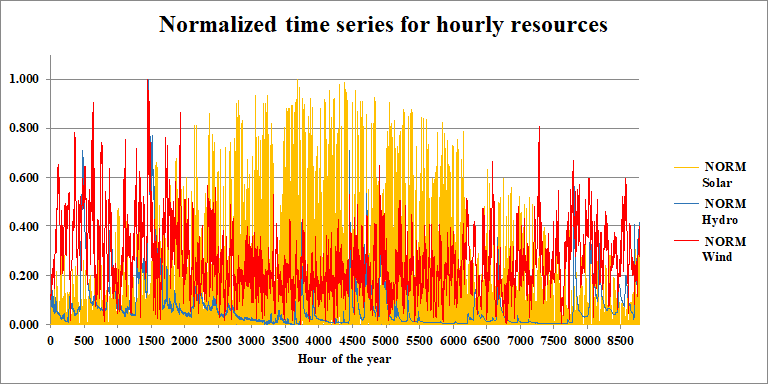}
    \caption{Normalized time series for hourly resources.}
    \label{fig:8}
\end{figure}
\begin{figure}
    \centering
    \includegraphics[width=\textwidth]{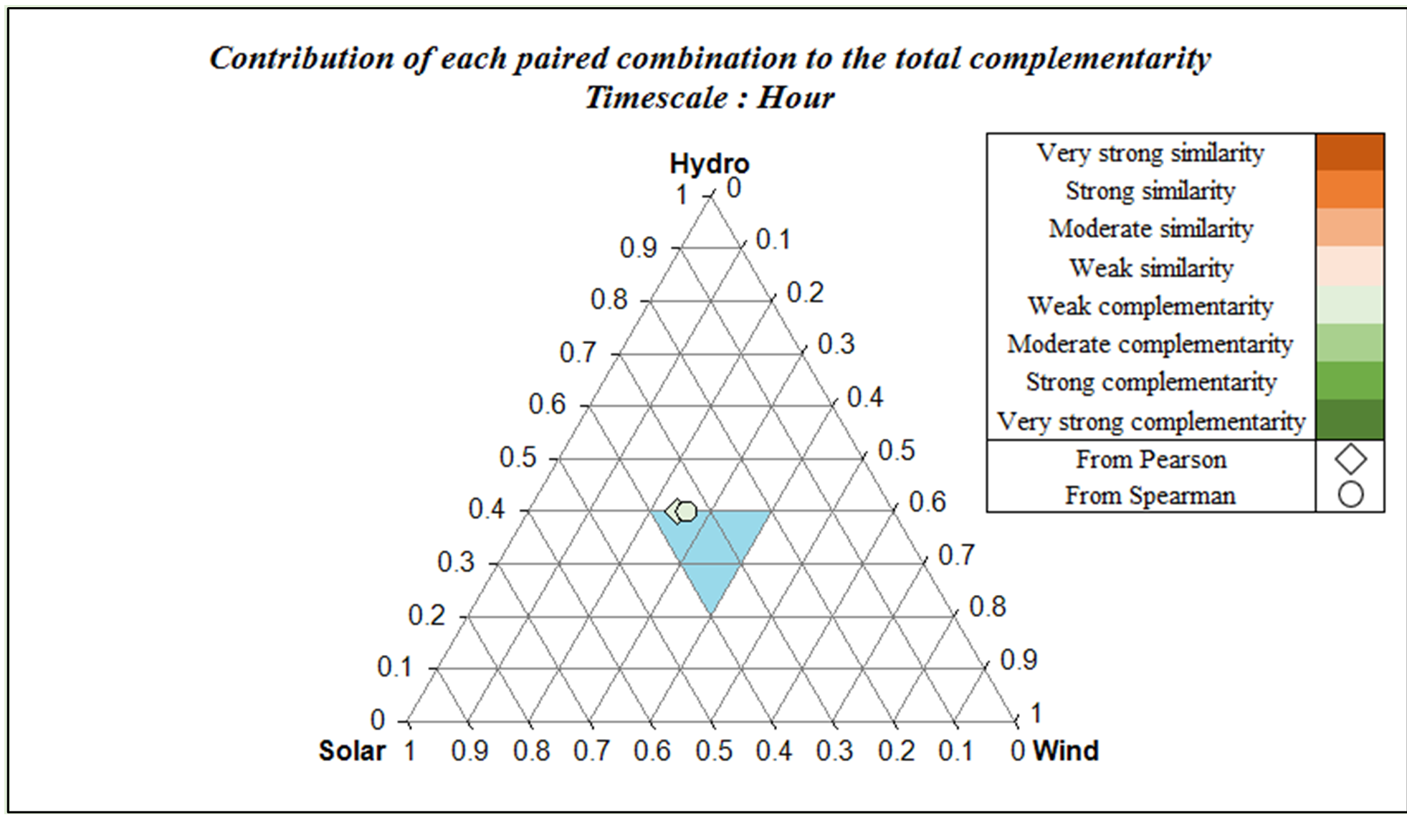}
    \caption{Contribution of each paired combination of VRES to $\kappa_t$ on a
hourly scale.}
    \label{fig:9}
\end{figure}
To illustrate the method presented in this paper, the three renewable resources available at the study area are evaluated on
three different timescales to assess energetic complementarity.
As in the method explanation, the indices for identifying each
resource are: w for wind, h for hydro and s for solar.
For assessing monthly complementarity between the three
sources, there were calculated the average hourly wind speed
and mean flow rate, as well as the mean daily sum of solar irradiation. Table 2 details the case study data and results for energetic complementarity, considering a monthly timescale.
These results show that on a monthly scale, the complementarity between wind and solar resources is the highest among
the three paired combinations, indicating that months with a
lower average solar irradiation are usually the ones with a
higher average windspeed, and vice versa. This is in line with the
findings presented by Jurasz et al. \cite{jurasz2018impact}. Similarly, the paired combination with the lowest complementarity (it presents a strong
similarity), corresponds to the combination of wind and hydro
resources ($\rho_{xy}$ = +0.717 - $\rho_s$ = +0.650), implying that both resources follow similar patterns (in terms of their lower and
higher values) along the year.

For the case study data, the application of the proposed
method results in a total monthly complementarity index $\kappa_t$
equal to 0.780 when considering Pearson Correlation Coefficient, and 0.830 for Spearman’s Rank. Therefore, and based on
Table 1, the joint behaviour between the three energy sources
on a monthly scale suggest a moderate to strong bundle complementarity. The ternary plot in Figure 5 displays the contribution of each paired combination to $\kappa_t$ , based on the normalisation of the individual distances calculated by means of equation
3. Using as illustrative example the results shown in the Figure
5, and based on $\rho_s$ , it is observed that wind-solar accounts for
50\% of $\kappa_t$ , whilst solar-hydro and wind-hydro contribute with
41\% and 9\%, correspondingly. The coloured triangle in the centre of the chart is the area where balanced distributions in terms
of contribution to $\kappa_t$ , would be plotted (for example, a perfectly
distributed contribution of the three VRES would be in the exact
centre of the plot).

The Figure 6 displays the normalized time series for the average daily resources. From this graph it is easily observed that
some complementarity exists between wind and solar, whilst
wind and hydro resources exhibit some similarity in their time
series on a daily scale. These observations are supported by the
results presented in Table 3, which shows that individual complementarities range from weak to moderate complementarity
or similarity.

Nevertheless, for this timescale, the results for $\kappa_t$ implies
that a hybrid power system might yield better results by considering a combination of these three resources, instead of a combination of just two of them, which would probably require a
significant amount of backup power or energy storage. The ternary plot in Figure 7 displays the contribution of each paired
combination to $\kappa_t$, and it can be mentioned that solar-hydro and
solar-wind resources have a similar impact on the value of $\kappa_t$,
suggesting that any hybrid system planning and operation
based on a daily scale should considered solar as the main energy source.

The normalized time series for the hourly resources is
shown in Figure 8, and the complementarity results are presented in Table 4. As expected, wind-solar complementarity is
heavily affected by the diurnal cycle, as was also shown by Jurasz et al. \cite{jurasz2018impact}, so much that the linear relationship between these
two resources is almost inexistent (Pearson presents a weak
complementarity, whilst Spearman’s rank correlation results in
a weak similarity). The correlation coefficients also evinced
weak similarity (wind-hydro) and weak complementarity (solar-
hydro) for the other paired combinations. For the bundle of the
three sources, the total temporal complementarity in an hourly
scale results in a $\kappa_t$ value of around 60\% of the maximum achievable, denoting an overall weak complementarity according to
Table 1. The contribution of each paired combination to $\kappa_t$ is
shown in Figure 9.

Based on the previous observations, some advantages of
the method become clear by comparing the results for complementarity between the three sources on each timescale. One of
them relates to its potential as a tool for preliminary assessment
of power generation planning and scheduling. The optimal
scheduling of systems that include a large renewables fraction
heavily depends on the forecasting accuracy of the correspond-
ing resources, which is in turn directly associated with the time-scale used for scheduling.

The method presented in this paper allows not only to assess the total complementarity between the three sources, but
also to evaluate the complementarity between each paired
combination of resources. Based on the results of the case
study, a system whose operation is based on a month-ahead
scheduling might be more effective by prioritising wind turbines
and photovoltaic panels along with their corresponding energy
storage devices. Similarly, a system in this region whose operation is based on a short-time or real time scheduling would
probably have a better performance by relying on a combination of the three VRES under consideration.
Correlation coefficients, compromise programming and normalisation are three proven mathematical methods, which
combined in the proposed method provide an interesting tool
for comparing different energy sources with different units of
measurement. Another advantage of this process is that it
makes the process versatile, because it admits useful modifications. For example, instead of assessing the resources complementarity, the complementarity could be assessed in terms of
power outputs by considering available head and efficiency (for
hydropower), temperature and PV panel characteristics (for solar power) and wind turbine specifications (for wind power).
Other variations could include capacity factors, blackout probability, etc. Another advantage of the method is that all results
are on a linear scale, providing a direct interpretation. Hybrid power systems set in areas with a low total temporal
complementarity index $\kappa_t$ , would benefit from hydropower reservoirs or other energy storage devices to improve their performance and reliability, whilst also avoiding the oversizing of the
system.
\section{Conclusions}
This paper presented a method for assessing energetic complementarity between three energy sources by means of a combination of correlation coefficients, compromise programming
and normalisation. This approach allows the simultaneous evaluation of partial and total complementarity. The correlation coefficients allow measuring the complementarity of each paired
combination of resources, whilst normalisation of compromise
programming results accounts for the total complementarity.
The method can be used in the design stages of autonomous
hybrid power systems. 

The case study results indicate that timescale selection has
a direct impact on the value of the vector components (and
hence on total complementarity). Therefore, this dependence
must be taken into consideration when using energetic complementarity indices for power generation planning and scheduling
purposes.

Correlation coefficients make possible the comparison of
different energy resources with different units of measurement.
In addition, the method allows practical modifications to assess
other complementarity features like power output, capacity
factors, etc. The linear scale and straightforward interpretation
of the results is another benefit of using the approach proposed
in this paper.

Based on the findings of this study and the previous paragraphs, future research might extend in the following directions:
\begin{itemize}
    \item Employing the method to assess spatial and spatiotemporal
energetic complementarity. This could also include the appraisal of complementarity between existing VRES plants.
\item Using the complementarity index $\kappa_t$ as a parameter or as part
of the objective function in an optimisation model, aiming to
define the best energy mix, improving reliability of the system
or determining the optimal operation schedule.
\item Including more than three variables energy sources and evaluate the impact on energy storage and capacity factors.
\end{itemize}

\section*{Acknowledgements}
The authors are grateful for the support received by their institutions for the research work that resulted in this paper. The
third author acknowledges the financial support received from
CNPq for his research work (proc. n.312941/2017-0).
\bibliographystyle{unsrt}  
\bibliography{references}  


\appendix
\section{Appendix - Minimum compromise programming
distance.}
\begin{figure}[!h]
    \centering
    \includegraphics[width=\textwidth]{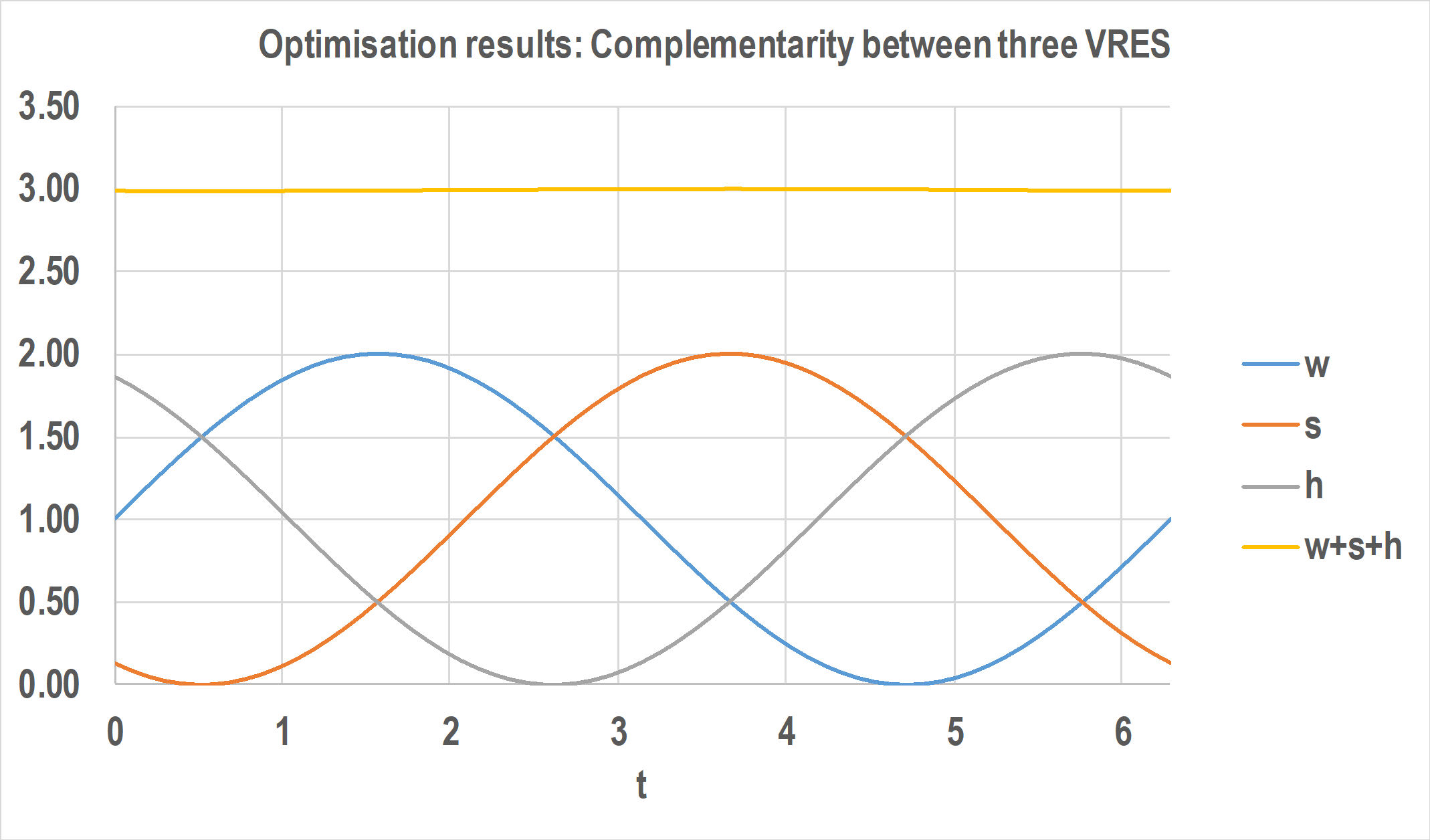}
    \caption{Normalized time series for hourly resources.}
    \label{fig:ten}
\end{figure}
Let us consider three idealised time series for w - wind, s - solar
and h - hydropower resources, each one of them having a sine
function form whose value depends on time - $t$ and phase shift
- $\Phi$:
\begin{align*}
    w(t) &= 1 + \sin(t)\\
    s(t) &= 1+sin(t+\Phi)\\
    h(t) &= 1+sin(t+2\Phi)
\end{align*}
The three paired combinations allow calculating three possible correlations coefficients – CC (Pearson or Spearman),
within the interval [-1, 1]. We can consider each correlation as
part of a three-dimensional vector c:
\begin{align}
    c(t) &= (CC_{ws},CC_{wh},CC_{sh})
\end{align}
For this system, the optimisation problem can be defined as
finding the $\Phi$ value that minimizes the compromise program-
ming distance, or $L_p$ metric:
\begin{align}
    \min L_p(c) &= \left[\sum_{k=1}^n \alpha_k^p \left|\frac{f_k^\text{best} - f_k(c)}{f_k^\text{best} - f_k^\text{worst}}\right|^p\right]^{\frac{1}{p}}
\end{align}
with:
\begin{itemize}
    \item $\alpha_k^p = 1$ for all cases;
    \item $p = 1$, direct proportion of magnitude
    \item $f_k(c) = CC$ value for paired combination $k$ of vector $c$;
    \item $f_k^\text{best} = -1$, representing full complementarity;
    \item $f_k^\text{worst} = 1$, representing full similarity;
    \item $0 \leq t,\Phi \leq 2 \pi$.
\end{itemize}
Satisfying the time interval constraints, it was found that
the $\Phi$ value ($\frac{2\pi}{3}$) results in a minimum $L_p(c) =
0.750$ with $CC = -0.5$ for the three paired combinations. Figure 10 displays the behaviour of the idealised time series when $\Phi = \frac{2\pi}{3}$.
As a general demonstration of this minimum achievable
$L_p(c)$ value, imagine we have n random variables $x_i$ , each with
unit variance, as in the case of correlation coefficients $\rho$.
Therefore, the sum of the variances:
\begin{align*}
    \text{Var} \left(\sum_{i=1}^n x_i\right) &= \sum_{i,j=1}^n \text{Cov}(x_i,x_j) \\
    &= \sum_{i=1}^n \text{Var}(x_i) + \sum_{i\neq j} \text{Cov}(x_i,x_j) \\
    &= n + \sum_{i \neq j} \rho_{ij} \\
    &= n + \frac{2n!}{2(n-2)!}\bar{\rho} \\
    &= n + n(n-1) \bar{\rho}
\end{align*}
In addition, we have $\text{Var} \left(\sum_{i=1}^n x_i\right) \geq 0$, that leads to:
\begin{align*}
\bar{\rho} \geq -\frac{1}{n-1}
\end{align*}
In consequence, for $n=3$, it results in $\bar{\rho} \geq -0.5$, therefore $L_p(c) \geq 0.75$.

\end{document}